\documentclass[preprint,12pt,authoryear]{elsarticle}

\usepackage{graphicx}
\usepackage{url}
\usepackage{apalike}

\usepackage{scrextend} 

\renewcommand{\figurename}{Fig.}

\journal{arxiv.org}

\begin{document}
\begin{frontmatter}

\title{Dopamine modulation via memristive schematic}

\author[inst1]{Max Talanov}
\ead{max.talanov@gmail.com}
\author[inst1]{Evgenii Zykov}
\ead{evgeniy.zykov@kpfu.ru}
\author[inst1]{Yuriy Gerasimov}
\ead{yurger2009@gmail.com}
\author[inst1]{Alexander Toschev}
\ead{atoschev@kpfu.ru}
\author[inst1,inst2]{Victor Erokhin}
\ead{victor.erokhin@fis.unipr.it}

\address[inst1]{KFU, Russia.}
\address[inst2]{CNR-IMEM, Italy.}


\begin{abstract}

  In this technical report we present novel results of the dopamine neuromodulation inspired modulation of a polyaniline (PANI) memristive device excitatory learning STDP. Results presented in this work are of two experiments setup computer simulation and physical prototype experiments. We present physical prototype of inhibitory learning or iSTDP as well as the results of iSTDP learning.  
  
\end{abstract}
\end{frontmatter}

\section{The experimental setup}\label{sec:experimental_setup}

\subsection{Block diagram}

\begin{figure}[ht]
  \centering
\includegraphics[width=1.0\textwidth]{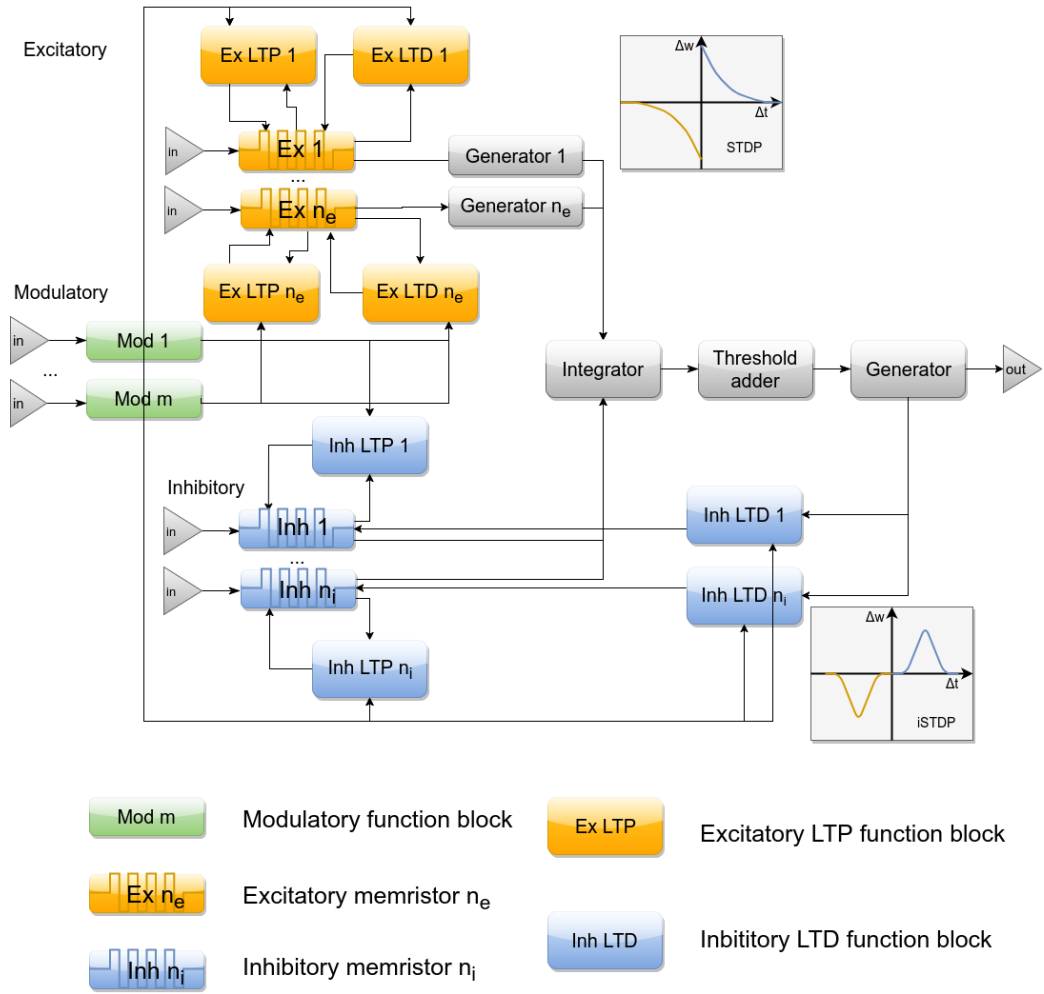}
\caption{Block diagram of modulatory, excitatory and inhibitory memristive neuron device. STDP and iSTDP graphs are used from \cite{vogels_inhibitory_2013,hennequin_inhibitory_2017}.\label{fig:block}}
\end{figure}

The block diagram is presented in the Figure \ref{fig:block} and is the memristive electronic implementation of an excitatory, inhibitory, neuromodulationary artificial neuron. There are three parts of the block diagram represented in colors: excitatory -- orange, inhibitory -- blue, modulatory -- green. Inputs are depicted as triangles. The excitatory learning is implemented via $Ex LTP [1..n_e]$ and $Ex LTD [1..n_e]$ feedback loops of excitatory memristive device ($Ex[1..n_e]$), where LTP (long term potentiation) blocks implement the learning function above $x$ axis and LTD (long term depression) blocks implement learning function below $x$. $Generator 1$ and $Generator n_e$ implement dendrite spikes. The inhibitory part has different structure, where the LTP is implemented as feedback loop block that uses inbound pre-synaptic signals and outbound of memristive device signals. The LTD is implemented as feedback from neuron outbound signal. The LTP and the LTD implement learning function described in \cite{hennequin_inhibitory_2017} indicated as iSTDP graph in Figure \ref{fig:block}. Modulatory blocks $Mod[1.. m]$ influence LTP and LTD functional blocks modulating the amplitude of learning impulses. $Integrator, Threshold adder$ and $Generator$ are the implementation of a neuronal soma and axon hillock that integrates excitatory and inhibitory inbound signals and generates the outbound signal.

\subsection{Wiring schematic}

\begin{figure}[ht]
  \centering
\includegraphics[width=1.5\textwidth,angle=90]{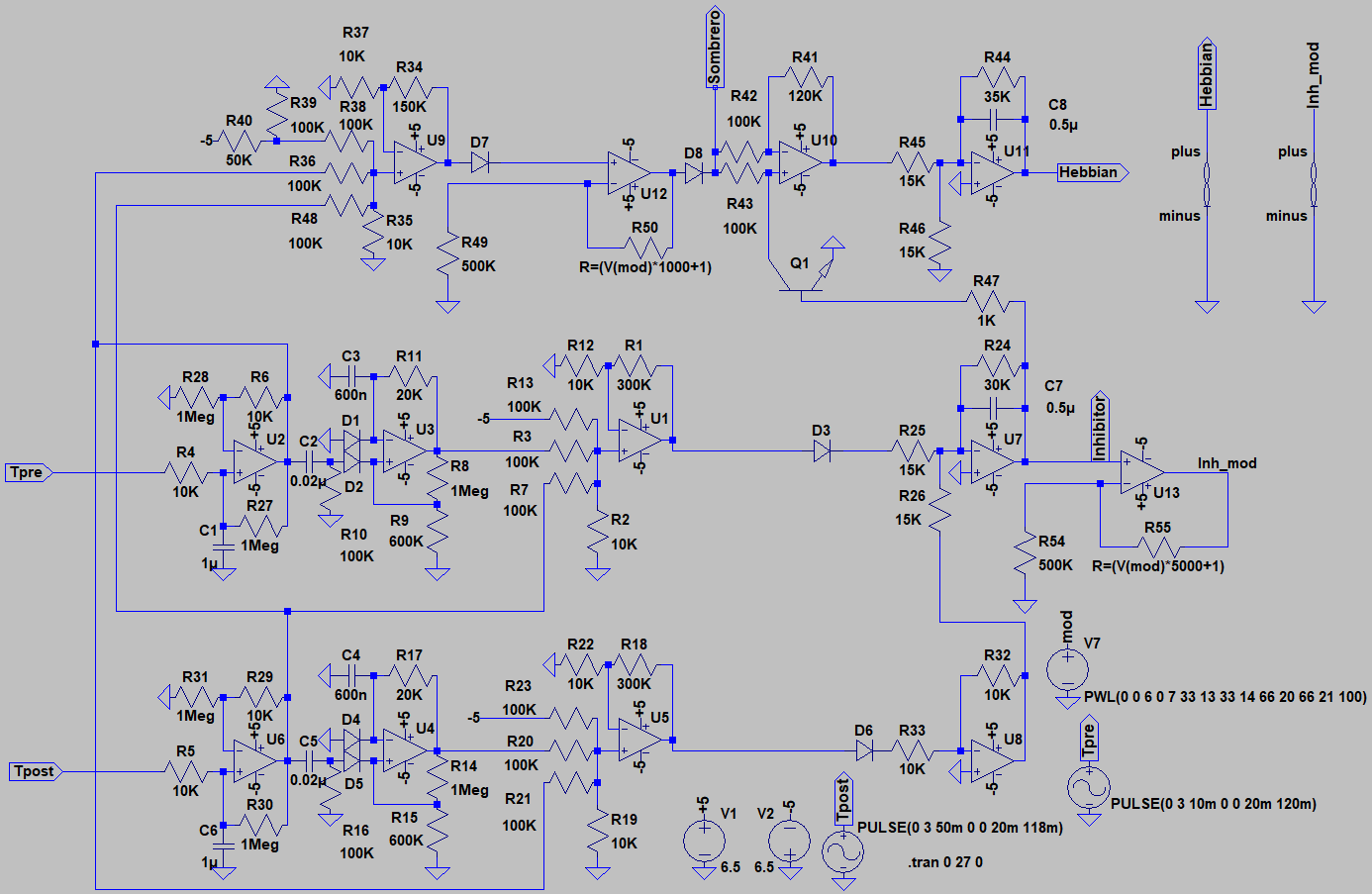}
\caption{Wiring schematic of modulatory, excitatory and inhibitory memristive neuron device.\label{fig:wiring}}
\end{figure}

\begin{figure}[ht]
\centering
\includegraphics[width=1\textwidth]{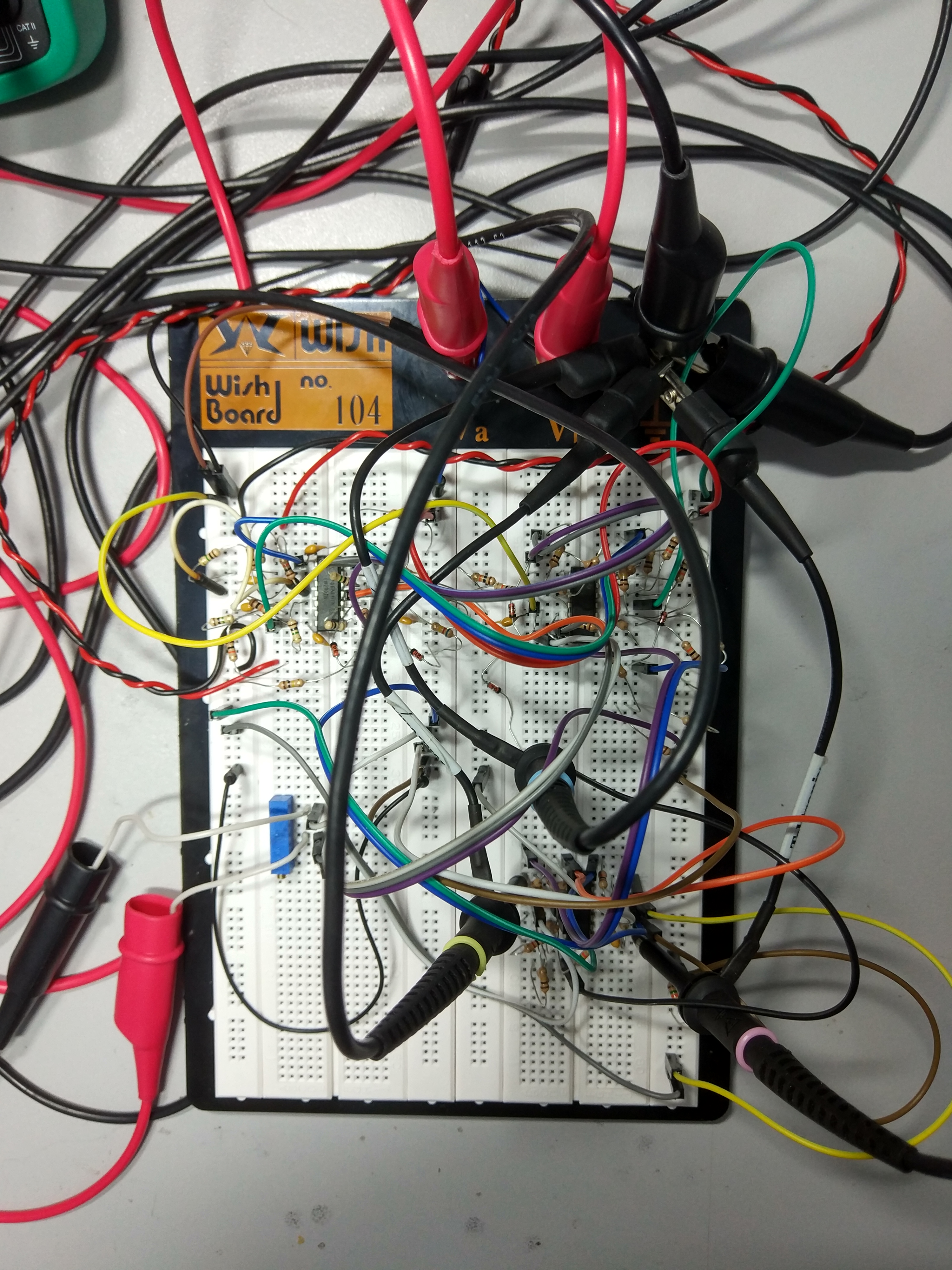}
\caption{Physical prototype implementation.\label{fig:physical_prototype}}
\end{figure}

The Figure \ref{fig:wiring} represents the wiring schematic, where excitatory and inhibitory learning impulses are transmitted to memristive elements.
Instead of the generator post-synaptic signals from Figure \ref{fig:block} we used an external generator  for simplicity of modeling.
The Hebbian STDP is implemented via op-amps $U9 –- U11$ and the iSTDP -- via op-amps $U1$, $U7$, $U8$.
Signals from $Tpre$ pre-synaptic spike generator input are transmitted to integrators implemented via  op-amp $U2$, which set the impulse descending edge of the learning function. The pulse-rise time constant of the integrating circuit is  $t = R4 \times C1$. When the accumulated voltage on the memristive elements exceeds the threshold, the one short multivibrator implemented via the operational amplifier $U3$ provides a single short pulse, which duration is determined by $T1 = C3 \times R11 \times \ln{(1 +\frac{R8}{R9})}$. Output signals from multivibrator are transmitted to the inverting adder implemented via $U1$.
Similarly, post-synaptic pulses from input $Tpost$ are created via on op-amps $U6$, $U4$ and $U5$ and then inverted via the op-amp $U8$. Output signals from both integrators are transmitted to the  adder-integrator op-amp $U7$ from which transmitted to inhibitory output $Inhibitor$. Signals from integrators $U2$ and $U6$ also are transmitted to the adder implemented via op-amp $U9$ and later transmitted to the controlled inverter the op-amp $U10$. When the non-inverting input of the operational amplifier the op-amp $U10$ is shorted to the ground, the operational amplifier works as an inverter; otherwise, it acts as a normal amplifier. Output positive pulse from $U7$ is applied to the key $Q1$ that controls a state of not inverting input of the controlled inverter of op-amp $U10$. From the output integrator on the op-amp $U11$ the signal is transmitted to excitatory $Hebbian$ output. The modulation of Hebbian STDP is preformed by op-amps $U12$ and $U13$. The function of the''bell'' form is implemented on the alternative output ''Sombrero''. The physical implementation of the wiring schematic is depicted in Figure \ref{fig:physical_prototype}.

\section{Results} \label{sec:results}

\subsection{Simulation}

\begin{figure}[ht]
\centering
\includegraphics[width=1.0\textwidth]{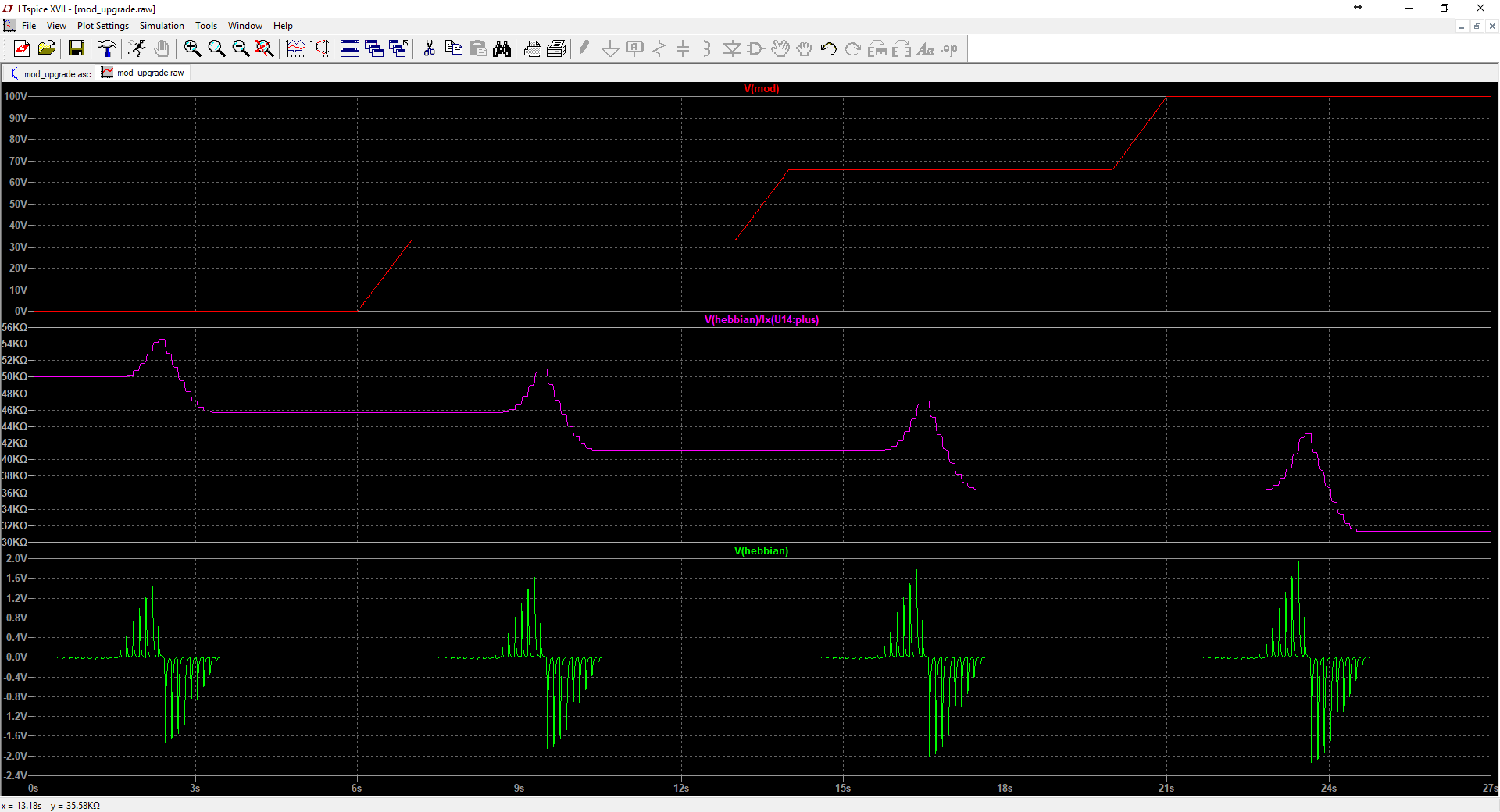}
\caption{The simulation results of learning (STDP):
  \emph{top} -- level of DA influence or setup of DA potentiometer ($V(mod)$ in Fig. \ref{fig:wiring}),
  \emph{middle} -- graph of memristive device conductivity,
  \emph{bottom} -- learning impulses (STDP). \label{fig:simulation}}
\end{figure}

The simulation results are presented in Figure \ref{fig:simulation}. The top graph depicts the level of the dopamine (DA) and identifies the level of modulation of learning impulses that is visible as the increment of green graph amplitude in the bottom graph, that in its turn influences the memristive device conductivity, described below. In the middle the lilac graph represents the result of the memristive device learning the overall conductivity. It is set by modulated learning impulses that are formed as Hebbian learning: $\Delta w = \frac{1}{\Delta t}$ where $\Delta t$ is the time lag between pre-synaptic spike and post-synaptic spike or inbound and outbound impulses. Pre-synaptic and post-synaptic spikes are presented in the Figure \ref{fig:wiring} as generators $Tpre$ and $Tpost$. For the simplification of the simulation purposes we used 2 different generators with phase shift to simulate different $\Delta t$s. This way we could depict whole Hebbian learning in one graph. Learning impulses are presented as bottom green graph in the Figure \ref{fig:simulation}. 

\subsection{Physical implementation}

\begin{figure}[t]
\centering
\includegraphics[width=0.45\textwidth]{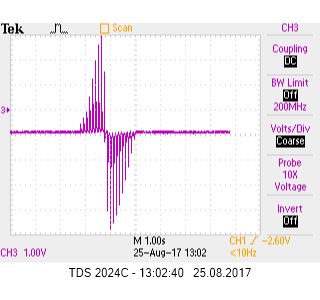}
\includegraphics[width=0.45\textwidth]{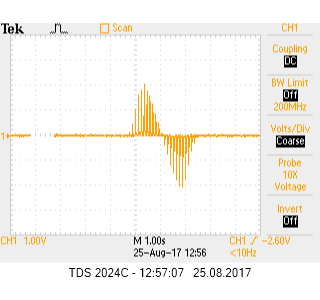}
\caption{The physical implementation, learning: \emph{left} -- STDP(Hebbian), \emph{right} -- iSTDP \label{fig:stdp_istdp}}
\end{figure}

Firstly we have implemented the learning functions for excitatory and inhibitory synapses, results are presented in the Figure \ref{fig:stdp_istdp} left is Hebbian STDP, right one is iSTDP as it was described in \cite{hennequin_inhibitory_2017} and presented in Figure \ref{fig:block}.

\begin{figure}[t]
\centering
\includegraphics[width=0.45\textwidth]{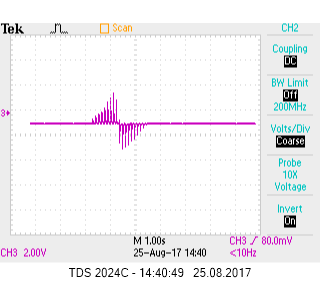}
\includegraphics[width=0.45\textwidth]{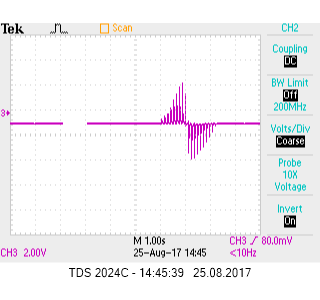}
\includegraphics[width=0.45\textwidth]{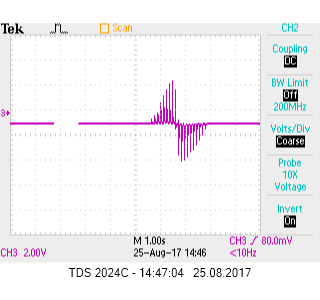}
\includegraphics[width=0.45\textwidth]{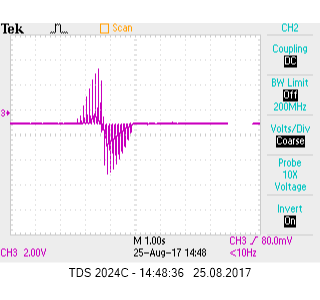}
\caption{The physical implementation DA modulation of the glutamate STDP \cite{gurney_new_2015}: \emph{top-left} -- $0/50 k\Omega$, \emph{top-right} -- $25/25 k\Omega$, \emph{bottom-left} -- $37.5/12.5 k\Omega$, \emph{bottom-right} -- $50/0 k\Omega$
  \label{fig:da_mod_stdp}}
\end{figure}

Second series of experiments was dedicated to the re-implementation of DA modulation of excitatory synapses STDP described in \cite{gurney_new_2015}. We have re-implemented biologically inspired modulatory function of DA. Results are depicted in the Figure \ref{fig:da_mod_stdp}, DA modulation is implemented as potentiometer $V(mod)$ presented in the Figure \ref{fig:wiring}. The top-left graph depicts learning impulses modulated with minimal level of DA potentiometer $0/50 k\Omega$, top-right is modulated by DA potentiometer $25/25 k\Omega$, bottom-left -- $37.5/12.5 k\Omega$, bottom-right -- $50/0 k\Omega$. The amplitude of learning impulses increases along with the modulation via DA potentiometer. 

Third series of experiments implements the DA modulation of iSTDP, that we assume should be similar to the DA modulation of STDP \cite{gurney_new_2015}, and this is just our assumption at the moment, as we could not find medical literature describing the DA modulation of inhibitory (GABA) synapses and requires further research. The top-left graph represents the iSTDP under influence of minimal DA modulatory potentiometer setup $0/1 M\Omega$, top-right -- $250/750 k\Omega$, bottom-left -- $500/500 k\Omega$ and bottom-right -- $750/250 k\Omega$. Again we indicate the increase of modulated amplitude of learning impulses along with increase of DA modulatory influence. 

\begin{figure}[t]
\centering
\includegraphics[width=0.45\textwidth]{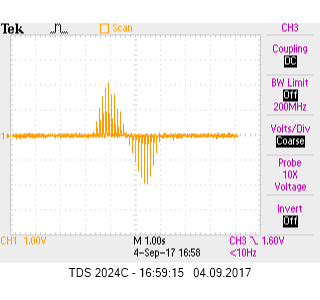}
\includegraphics[width=0.45\textwidth]{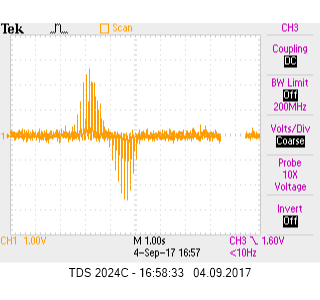}
\includegraphics[width=0.45\textwidth]{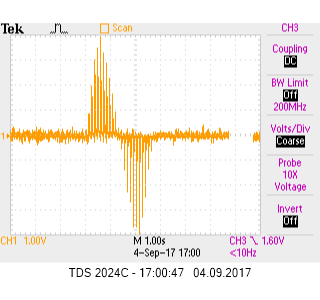}
\includegraphics[width=0.45\textwidth]{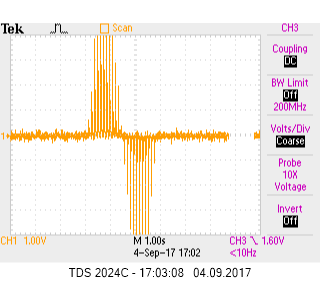}
\caption{The physical implementation DA modulation of the iSTDP: \emph{top-left} -- $0/1 M\Omega$, \emph{top-right} -- $250/750 k\Omega$, \emph{bottom-left} -- $500/500 k\Omega$, \emph{bottom-right} -- $750/250 k\Omega$
  \label{pic:da_mod_istdp}}
\end{figure}

\bibliographystyle{apalike}
\bibliography{bib2017}
\end{document}